\documentclass{PoS}

\usepackage{subfig}
\usepackage{url}
\usepackage{cite}
\newcommand{\ncteq}{{\tt nCTEQ}}
\newcommand{\ncteqfit}{{\tt nCTEQ15}}
\newcommand{\ncteqnp}{{\tt nCTEQ15-np}}

\title{Nuclear parton distributions from the nCTEQ group}

\ShortTitle{nPDFs from the nCTEQ group}

\author{\speaker{A. Kusina}\\ 
        Laboratoire de Physique Subatomique et de Cosmologie,\\
              Universit\'e Grenoble-Alpes, CNRS/IN2P3,\\
              53 avenue des Martyrs, 38026 Grenoble, France\\
        E-mail: \email{kusina@lpsc.in2p3.fr}}

\abstract{We present the nCTEQ15 global analysis of nuclear parton
distribution functions (nPDFs). The main addition to the
previous nCTEQ PDFs is the introduction of PDF uncertainties based on
the Hessian method. Another important improvement is the inclusion of
pion production data from RHIC giving us a handle to constrain gluon
PDF. In this presentation we briefly discuss the framework of our
analysis and concentrate on the comparison of our results with those
of other groups. Additionally we show a first estimate of the impact
of the LHC pPb $W/Z$ boson production data on the presented nCTEQ15 PDFs.}

\FullConference{XXIV International Workshop on Deep-Inelastic Scattering and Related Subjects\\
		11-15 April, 2016\\
		DESY Hamburg, Germany}

\begin{document}

\section{Introduction}
\label{intro}

Nucleons and nuclei can be described using the language of parton distribution
functions (PDFs) which is based on factorization
theorems~\cite{Collins:1985ue,Bodwin:1984hc,Qiu:2002mh}.
The case of a free proton is extremely  well studied. Several global analyses of
free proton PDFs, based on an ever growing set of precise experimental data and
on next-to-next-to-leading order theoretical predictions, are regularly updated
and maintained~\cite{Martin:2009iq,Gao:2013xoa,Ball:2012cx,Owens:2012bv}.
The structure of a nucleus can be effectively parametrized in terms of protons bound
inside a nucleus and described by nuclear PDFs (nPDFs). These nPDFs contain effects
on proton structure coming from the strong interactions between
the nucleons in a nucleus. Similarly to the PDFs of free protons, nuclear PDFs are
obtained by fitting experimental data including deep inelastic scattering on nuclei
and nuclear collision experiments. Moreover, as the nuclear effects are clearly dependent
on the number of nucleons, experimental data from scattering on multiple nuclei must
be considered. In contrast to the free proton PDFs where quark distributions for most
flavors together with the gluon distribution are reliably determined over a large
kinematic range,  nuclear PDFs precision is not comparable due to the lack of accuracy
of the current relevant data. In addition, the non-trivial dependence of nuclear effects
on the number of nucleons requires a large data set involving several different nuclei.
Nevertheless, nuclear PDFs are a crucial ingredient in predictions for high energy
collisions involving nuclear targets, such as the lead collisions performed at the LHC.

In this contribution we present the new {\tt nCTEQ15} nuclear PDFs that were recently
released~\cite{Kovarik:2015cma} and compare them with analyses from other groups providing
nPDFs~\cite{Hirai:2007sx,Eskola:2009uj,deFlorian:2011fp}. Additionally we present a first
estimate of the impact of the pPb LHC $W/Z$ boson production data on the presented
{\tt nCTEQ15} PDFs.

\section{\ncteq\ global analysis}
In the presented \ncteq\ analysis we use mostly charged lepton deep inelastic scattering
(DIS) and Drell-Yan process (DY) data that provide respectively 616 and 92 data points.
Additionally we include pion production data from RHIC (32 data points) that have potential
to constrain the gluon PDF. To better asses the impact of the pion data on our analysis
two fits are discussed: (i) the main \ncteqfit\ fit using all the aforementioned data,
and (ii) \ncteqnp\ fit which does not include the pion data.
The framework of the current analysis, including parameterization, fitting procedure and
precise prescription for the Hessian method used to estimate PDF uncertainties is
defined in ref.~\cite{Kovarik:2015cma} to which we refer the reader for details.

In both presented fits, we use 16 free parameters to describe the nPDFs, that comprise
7 gluon, 4 $u$-valence, 3 $d$-valence and 2 $\bar{d}+\bar{u}$ parameters.
In addition, in the \ncteqfit\ case the normalization of the pion data sets is fitted
which adds two more free parameters.
Both our fits, \ncteqfit\ and \ncteqnp\ describe the data very well.
Indeed, the quality of the fits as measured by the values of the $\chi^2/$dof
(0.81 and 0.84 for the \ncteqfit\ and \ncteqnp\ fits respectively), confirms it.
%
%
\begin{figure*}[th]
\centering{}
\subfloat[]{
\includegraphics[width=0.49\textwidth]{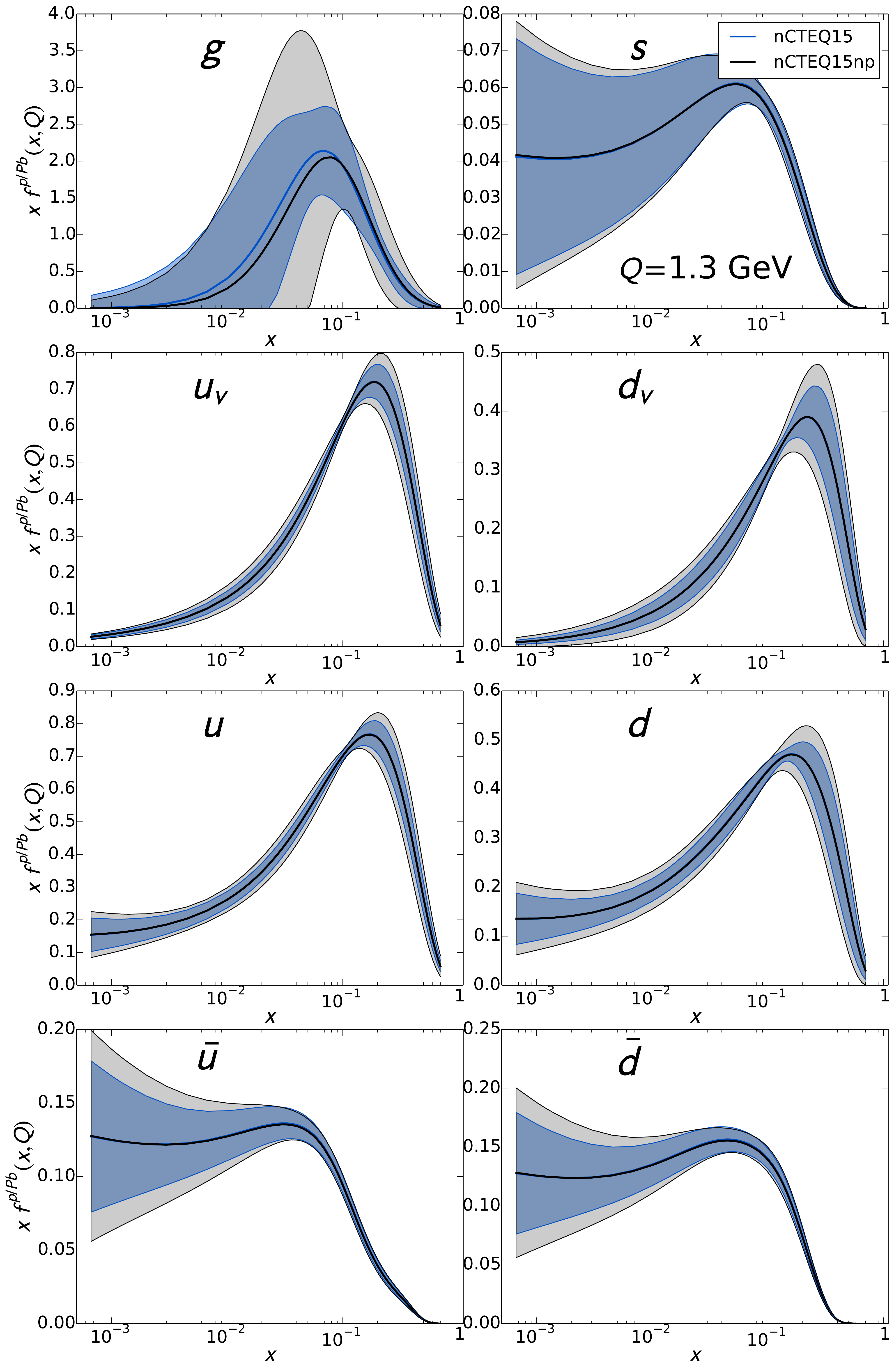}
\label{subfig:ncteq15-vs-ncteq15np}
}
\subfloat[]{
\includegraphics[width=0.48\textwidth]{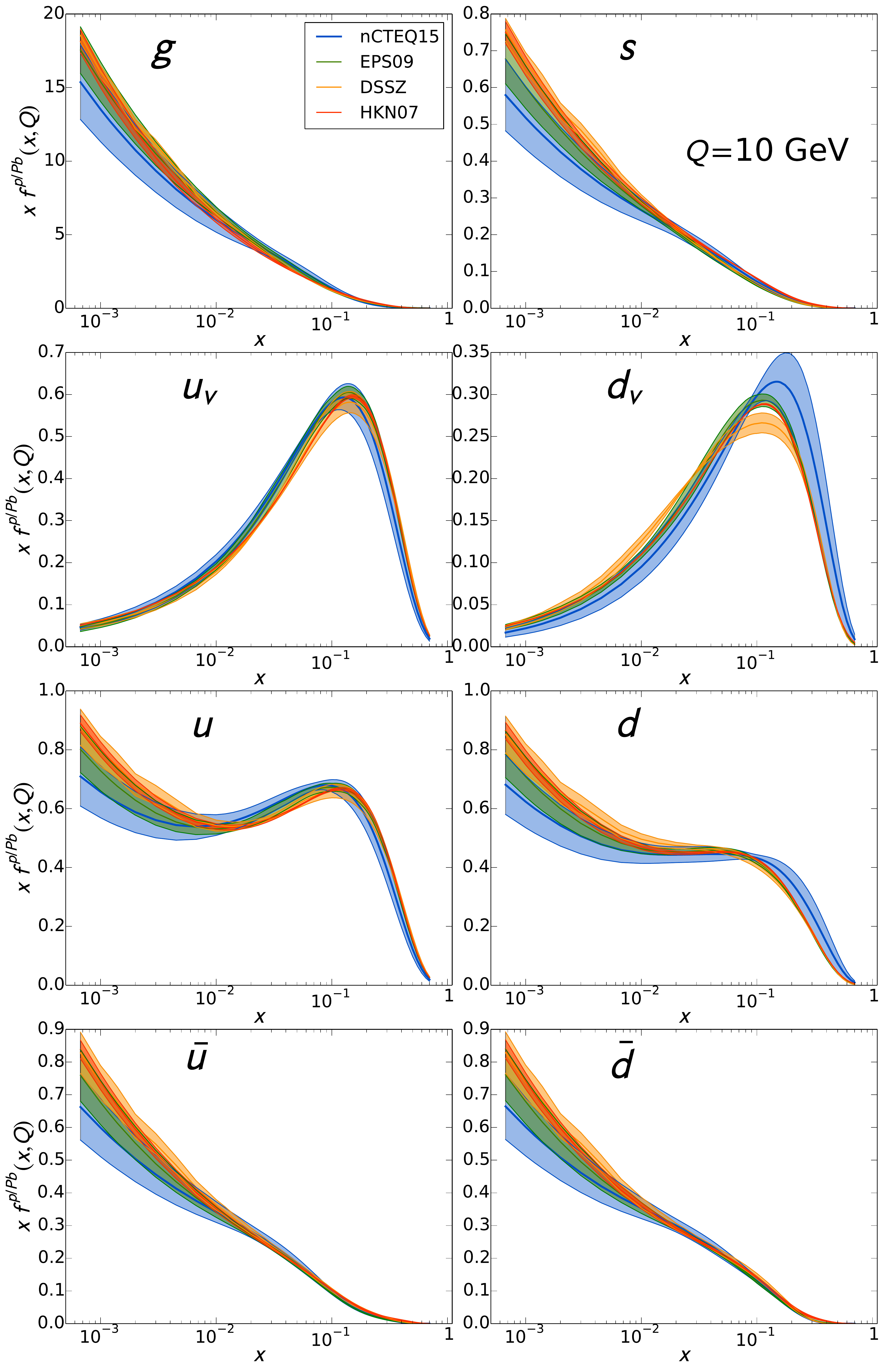}
\label{subfig:ncteq15-vs-others}
}
\caption{
(a) Comparison of bound proton lead PDFs from the \ncteqfit\ fit (blue)
and the \ncteqnp\ fit without pion data (gray) at the initial scale $Q=1.3$ GeV.
(b) Comparison of  the \ncteqfit\ fit (blue) with results from other groups:
EPS09 (green), DSSZ (orange), HKN07 (red). Shown are bound proton lead PDFs
at scale $Q=10$~GeV.}
\label{fig:ncteqPDFs}
\end{figure*}
%
Figure~\ref{subfig:ncteq15-vs-ncteq15np} shows the bound proton PDFs resulting from the two fits.
It clearly shows that the pion data impact
the gluon distribution, and to a lesser extent the $u_v$, $d_v$ and $s$ PDFs.
The inclusion of the pion data decreases the lead gluon PDF at larger $x$
($\gtrsim 10^{-1}$), and increases it at smaller $x$ whereas the error bands are
reduced in the intermediate to larger $x$ range.
For most of the other PDF flavors, the change in the central value is minimal.
For these other PDFs,
the inclusion of the pion data generally decreases the size of the error band.

\section{Comparison with other nPDFs}

We now compare the \ncteqfit\ PDFs with other recent nuclear parton distributions
in the literature, in particular 
HKN07~\cite{Hirai:2007sx},
EPS09~\cite{Eskola:2009uj}, and
DSSZ~\cite{deFlorian:2011fp}.
Our data set selection and technical aspects of our analysis are
closest to that of EPS09 on which we focus our comparison.
As an example in Fig.~\ref{subfig:ncteq15-vs-others} we present comparison of the
bound proton lead PDFs at the scale $Q=10$~GeV from the different groups.

For most flavors, $\bar{u}$, $\bar{d}$, $s$ and $g$, there is a
reasonable agreement between predictions.
Even though, for the gluon, there is a larger spread in the predictions form
the various PDF sets;
we can see a distinct shape predicted by the \ncteqfit\ and EPS09 fits
whereas HKN07 and DSSZ have similar, much flatter behavior in the small to
intermediate $x$ region and deviates from each other in the higher $x$ region;
however, all these differences are nearly contained within the PDF uncertainty bands.

Examining the $u$- and $d$-valence distributions, 
one can see that  HKN07, EPS09, DSSZ sets  
agree quite closely with each other throughout the $x$ range.
While the \ncteqfit\ fit uncertainty bands generally overlap with the other
sets, we see on average the $u_v$ distribution is softer while the $d_v$
distribution is harder.
This difference highlights an important feature of the \ncteqfit\ fit; namely,
that the $u_v$ and $d_v$ are allowed to be independent, whereas other groups assume
the corresponding nuclear corrections to be identical. 
There is no physical motivation to assume the $u_v$ and $d_v$ nuclear corrections
to be universal however the sensitivity of the currently available data to these differences
is limited,%
    \footnote{One of the reasons for this lack of sensitivity is the fact that older DIS
    data have been corrected for the neutron access and in turn have lost its ability to
    distinguish between $u_v$ and $d_v$ distributions.}
which allows for good data description even with this assumption.

\begin{figure*}[h!!!]
\centering{}
\includegraphics[clip=true,width=0.99\textwidth]{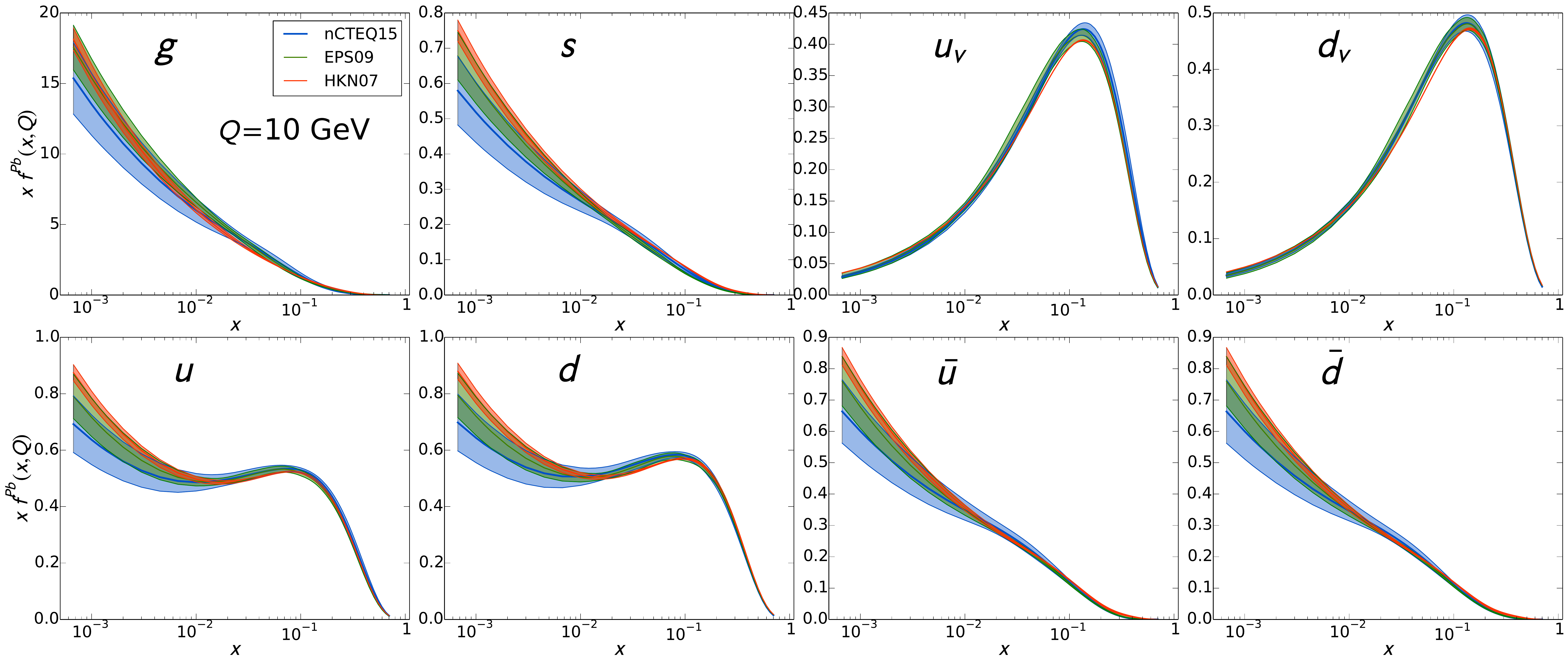}
\caption{
Full nuclear lead PDFs from different groups at the scale of $Q=10$ GeV.
}
\label{fig:full-nPDF}
\end{figure*}

This additional freedom in the \ncteqfit\ valence distributions results in the 
difference between the bound proton valence distributions that is seen in
Fig.~\ref{subfig:ncteq15-vs-others}. Even though the difference is substantial
we need to remember that the bound proton
distributions are not really objects of interest, they are merely a very convenient way of
parameterizing the actual quantities that are physically important -- the full nuclear PDFs.
The nuclear PDFs provide the distributions of partons in the whole nucleus and are combinations
of bound proton and bound neutron PDFs
\begin{equation}
f^{A}=Z/A f^{p/A} + (A-Z)/Z f^{n/A},
\end{equation} 
with $Z$ being the number of protons and $A$ the number of protons and neutrons in the nucleus.
If we examine the differences between the full nuclear PDFs of the different groups,
Fig.~\ref{fig:full-nPDF}, we can see that the agreement between valence distributions is excellent.
This means that the relatively big discrepancy on the level of bound proton valence PDFs vanishes
due to the averaging of $u$ and $d$ distributions occurring when bound proton and bound neutron PDFs
are summed.

\section{Impact of the LHC data}
\label{reweighting}
\begin{figure}[th]
\centering{}
\subfloat[$u$ distribution]{
\includegraphics[width=0.48\textwidth]{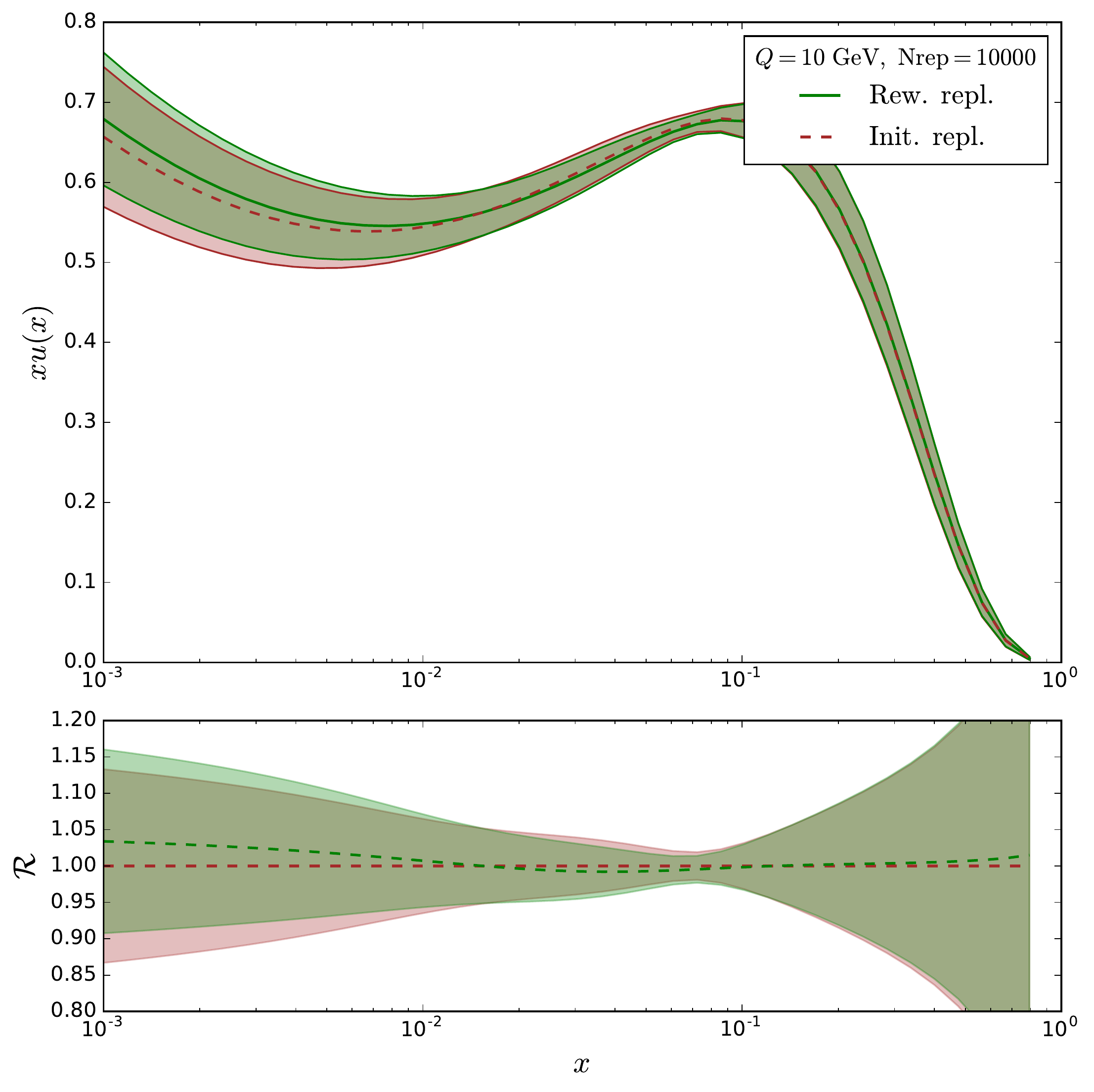}
}
\hfil
\subfloat[$\bar{d}$ distribution]{
\includegraphics[width=0.48\textwidth]{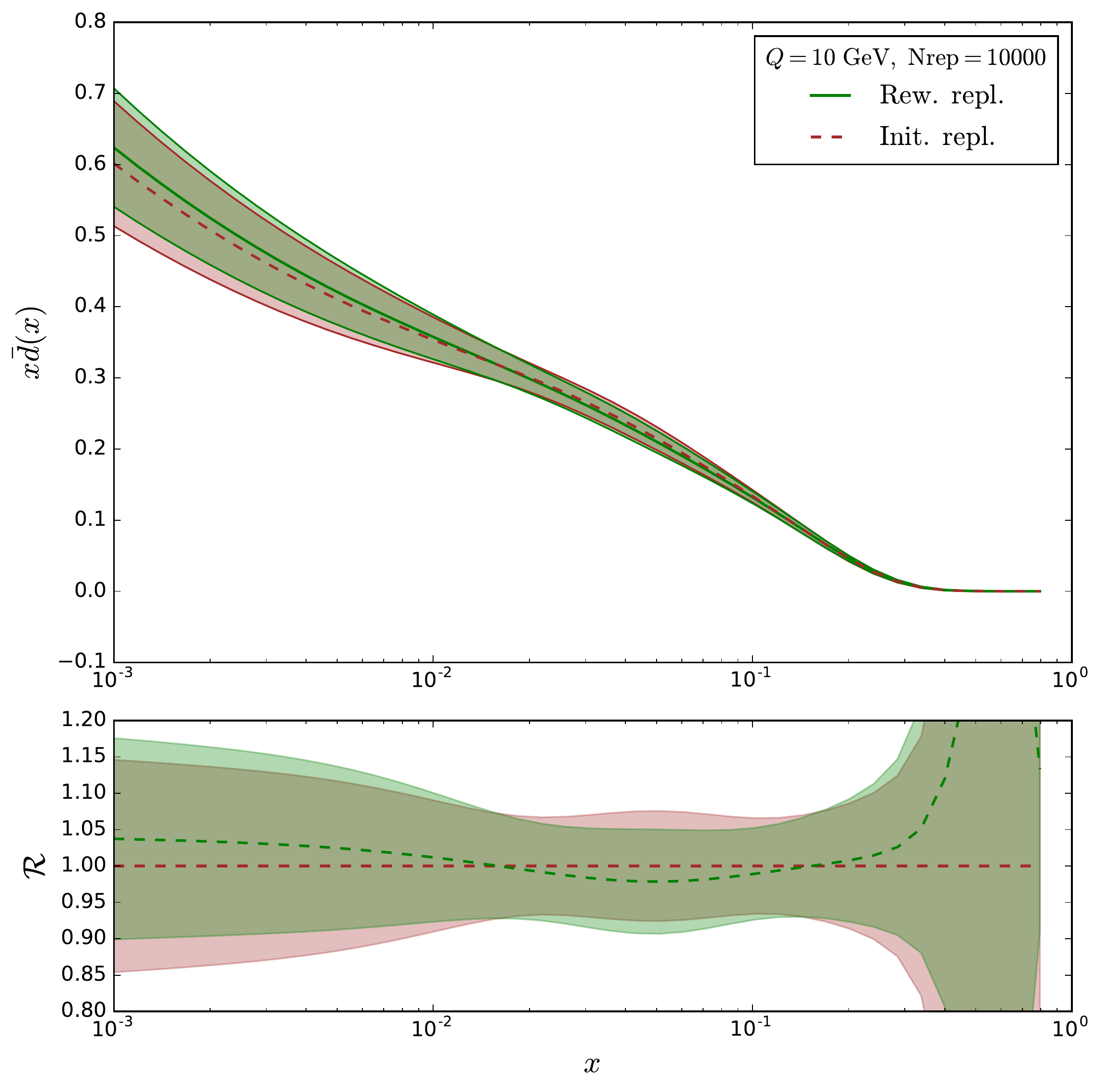}
}
\caption{Comparison of PDFs after and before the reweighting procedure.}
\label{fig:pdfs_after_rew}
\end{figure}

The LHC proton-lead runs provide data that can be very valuable for constraining the nPDFs.
One important example of such data is $W$ and $Z$ boson production.
We are currently using the already available vector boson
data~\cite{Aad:2015gta,AtlasWpPb2,Khachatryan:2015pzs,Khachatryan:2015hha,Aaij:2014pvu,Senosi:2015omk}
to estimate their impact on the \ncteqfit\ PDFs using the reweighting technique. We show here
preliminary results of this study the full results will be available soon~\cite{Kusina:2016fxy}.
To perform the reweighting we employ the original weight definition introduced in~\cite{Giele:1998gw},
additionally accounting for the tolerance criterion used in our fit as advocated
in~\cite{Paukkunen:2014zia,Armesto:2015lrg}.
In Fig.~\ref{fig:pdfs_after_rew} we show the effect of these data on PDFs by comparing $u$ and
$\bar{d}$ distributions before and after the reweighting. The effects are currently limited
mainly due to the rather large experimental uncertainties of the data. 
The dominant effect is coming from the CMS $W^{\pm}$ measurement which features the smallest errors.
Similar conclusions have been obtained by the EPS group~\cite{Armesto:2015lrg} where also jet data have
been used showing their potential for constraining the nuclear gluon.
More pPb data will be collected this year which should provide more stringent constraints
on the current nPDFs. Additionally, even with the current limitted precision of the pPb data
including them directly into the nPDF fits could allow to open new degrees of freedom (e.g. strange)
leading to more realistic nPDFs (this effects are, however, hard to estimate using the reweighting technique).

\section{Conclusions}

We have presented the recent \ncteqfit\ nuclear PDFs. The analysis have been performed in 
the {\tt CTEQ} framework and used Hessian method to determine PDF errors. The resulting nPDFs
are publicly available in our internal PDS format (with corresponding interface) as well
as in the new LHAPDF6 format. They can be downloaded from the \ncteq\ \cite{nCTEQwebpage}
and LHAPDF~\cite{LHAPDFwebpage} websites. 

We find relatively good agreement between our \ncteqfit\ nPDFs and those from other groups
especially with EPS09. However, there are certain differences in both methodologies
and results. One of them is the difference in treatment of the valence distributions which leads
to differences at the level of bound proton PDFs which, however, vanish when full nuclear PDFs
are constructed.

The errors of the \ncteqfit\ PDFs are comparable in size to those of EPS09 but they tend to
be bigger than the HKN and DSSZ ones. Even with these relative consistency in the error
determination it should be kept in mind that nPDF errors are still significantly underestimated.
This is caused by the limited number of free parameters in the fitting procedure and assumptions
like the one on the valence distributions. Unfortunately this kind of assumptions are currently
unavoidable due to the lack of experimental data covering different kinematic regions.

The LHC pPb (and to a lesser extent PbPb) data have the potential to help further constrain nPDFs
and in particular to obtain better sensitivity to the difference between $u_v$ and $d_v$
distributions. Unfortunately their current precision is still limited but more data will be
available this year.

\bibliographystyle{utphys_spires}
\bibliography{biblio}
%

\end{document}